\documentclass[a4paper,11pt]{article}
\pdfoutput=1 
\usepackage{amsmath}
\usepackage{amsthm}
\usepackage{braket}
\usepackage{jinstpub_prova} 

\usepackage{subfigure}

\title{\boldmath Performance of the BIS78 RPC detectors: a new concept of electronics and detector integration for high-rate and fast timing large size RPCs}

\author[a,b,1]{L. Pizzimento,}

\affiliation[a]{University of Rome Tor Vergata  }
\affiliation[b]{INFN Roma 2, Roma, Italia}

\emailAdd{luca.pizzimento@roma2.infn.it}

\abstract{The reduction of the average charge per count in the gas along with the capability to discriminate very small signals from noise can allow an efficient and long-term Resistive Plate Chamber detector operation in high radiation background environment. This goal has been reached during the R$\&$D program of the BIS78 upgrade project of the ATLAS detector at LHC through the deep integration of a fast (100 ps peaking time) and sensitive (as small as 100 $\mu$V threshold) Front-End electronics (FE) with a very large size detector structure. This innovative RPC integration concept pivots on a newly conceived faraday cage, embedding the readout strips and the FE, tightly wrapped around a 1 mm gas gap RPC with 1.2 mm thick electrodes, as a fully independent singlet structure. We studied the performance of BIS78 production triplet chambers, made of 3 independent singlets of 2 $m^{2}$, each providing a 2D space and time information, showing a minimum threshold achievable of 3 pC of average charge per count produced inside the gas gap. We show that these chambers grant a record combined performance of better than $95\%$ single gap efficiency, time resolution of 350 ps and  10 $kHz/cm^{2}$ rate capability.

 }


\collaboration[c]{on behalf of ATLAS collaboration}

\proceeding{XV RPC 2020 - Workshop on resistive plate chambers and related detectors\\
  10-14 February 2020\\
  Rome, ITALY}

\begin{document}
\maketitle
\flushbottom
\section{BIS78 ATLAS RPC}\label{intro}
The BIS78 project is the pilot project \cite{ATLASTDRMS} for the barrel inner (BI) plane of the muon spectrometer upgrade scheduled for the Phase-2 ATLAS upgrade. These chambers are designed to be compatible with the high luminosity LHC (HL-LHC) conditions and the BI RPCs will inherit most of the design. The BIS78 RPC chambers will be deployed in the Phase-1 upgrade, providing a new system which integrates small-diameter Muon Drift Tube (sMDT) + RPC detectors to be installed in the barrel-endcap transition region at $1.0 < |\eta| < 1.3$, which is the region characterized by the highest background rates in the barrel. The estimated total rate of the Level-1 single-muon trigger with $p_{T}> 20$ GeV is expected to rise during Run-3 up to $57.6$ $kHz$, while the allocable is $25$ $kHz$ for muon triggers out of a total bandwidth of $100$ $kHz$. In order to address this problem the pseudorapidity region $1.3 < |\eta| < 2.5$ will be covered by the New Small Wheels \cite{ATLASTDRMS} that will solve the problems of fake triggers mainly coming from secondary charged particles generated inside the cryostats and will improve the muon selectivity and the discrimination of low momentum muons. The region at $1.0 < |\eta| < 1.3$, which corresponds to the barrel-endcap transition region, will be covered with the BIS78 RPC chambers, leading to a total trigger coverage of $\approx83.5\%$ of the transition region.
The BIS78 stations are formed by two triplets of RPCs coupled to a multilayer sMDT in the same envelope, without making any electrical contact between the two systems preserving their electrical and mechanical integrity. The BIS78 RPC system consists of 16 BIS7 and 16 BIS8 triplets. The total area covered by this system is 94 $m^{2}$. 
The new RPCs have a gas gap width of 1 mm. This choice allows to reduce the total thickness of the system and leads to other benefits: the RPC time resolution scales with the gas gap width. Thus this reduction leads to an improvement of the time resolution up to $\approx 0.4$ $ns$. The resistive electrodes thickness also has been reduced to $1.2$ $mm$. This reduction takes the advantage of increasing the charge fraction transferred to the pick-up electrodes, resulting in an improvement of the Signal-to-Noise ratio. Each detector is readout on both sides by two panels of orthogonal strips, with strip pitches of $24$-$26$ $mm$ depending on the chamber type, providing $\eta$ and $\phi$ coordinates. The challenge of making the chambers compatible with operation at higher hit rates is addressed by a proportional reduction of charge per count, while simultaneously increasing the sensitivity and signal-to-noise ratio of the Front-End electronics without degrading the fast timing features of the RPCs. In order to improve the rate capability a new Front End electronics has been realized. Also a new Faraday cage design, suitable for low-threshold operation, is being developed, allowing a better shielding of the more sensitive Front-End electronics. All these inter-dependent requirements have led to a full re-design of the RPC, optimizing it in all aspects, from materials, to readout electronics, up to chamber layout. The new-generation RPCs will increase the rate capability by an order of magnitude, decrease the total chamber weight and thickness, operate at near half the working voltage. The main features of the BIS78-type RPCs compared to the ATLAS Standard ones are reported in Table \ref{BIS78recap}.
\begin{table}[h!]
\centering
\caption{RPC Main parameters comparison between ATLAS standard and BIS78 upgrade}

\begin{tabular}{||c | c | c ||} 
 \hline
 Detector parameters &ATLAS RPC&BIS78 RPC \\ [0.5ex] 
 \hline\hline
 Gas gap width &2 mm&1 mm\\ 
\hline 
 Electrode Thickness& 1.8 mm& 1.2 mm\\
\hline 
Time Resolution&$\approx1$ ns&$\approx 0.4$ ns\\
\hline  
Space Resolution&$\approx6$ mm&$\approx 1$ mm\\
\hline 
Gaps per chamber&2&3\\
\hline 
Gas Mixture&ATLAS Standard&ATLAS Standard\\
\hline 
Readout&2D Orthogonal&2D Orthogonal\\
\hline 
FE technology&GaAs&Si$\&$Si-Ge\\
\hline 
FE Effective Threshold&2-3 mV& 0.2-0.3 mV\\
\hline 
FE Power consumption&30 mW/ch& 12 mW/ch\\
\hline
\end{tabular}
\label{BIS78recap}
\end{table}

The RPC \cite{santonico}-\cite{cardarelli} rate capability is mainly limited by the current that can be driven by the high resistivity electrodes. It can be improved by modifying the highly interconnected parameters which define the voltage drop on the electrodes. These parameters may be derived by applying the Ohm's law:
\begin{equation}
V_{gas}=V_{a}-R \cdot I=V_{a}-\rho \cdot \dfrac{d}{S} \cdot \braket{Q} \cdot S \cdot \Phi_{particles}=V_{a}-\rho \cdot d \cdot \braket{Q} \cdot \Phi_{particles}
\end{equation}
Where $V_{gas}$ is the voltage applied across the gas, $V_{a}$ is the total voltage supplied to the detector, $\rho$ is the bulk resistivity of the electrodes, $d$ is the electrodes thickness, $\braket{Q}$ is the average charge per count produced inside the gas, $S$ is the detector area and $\Phi_{particles}$ represents the flux of ionizing particles. 
This equation shows several ways to increase the detectable particle flux, however, the approach chosen by this project is to reduce the average charge per count $\braket{Q}$, allowing to increase the rate capability while operating the detector at fixed current. The drawback of this approach is that reducing the average charge per count implies a reduction of the injected charge inside the Front End, hence the signal which needs to be detected will be much smaller. For this reason, a  very sensitive Front End electronics is mandatory in order to detect such small signal. Moreover, a high suppression of the noise induced inside the detector by the electronics and by external sources is required, leading to the chamber structure optimization as a Faraday cage.

\begin{center}
\begin{figure}[htbp]
\centering
\includegraphics[scale=0.35]{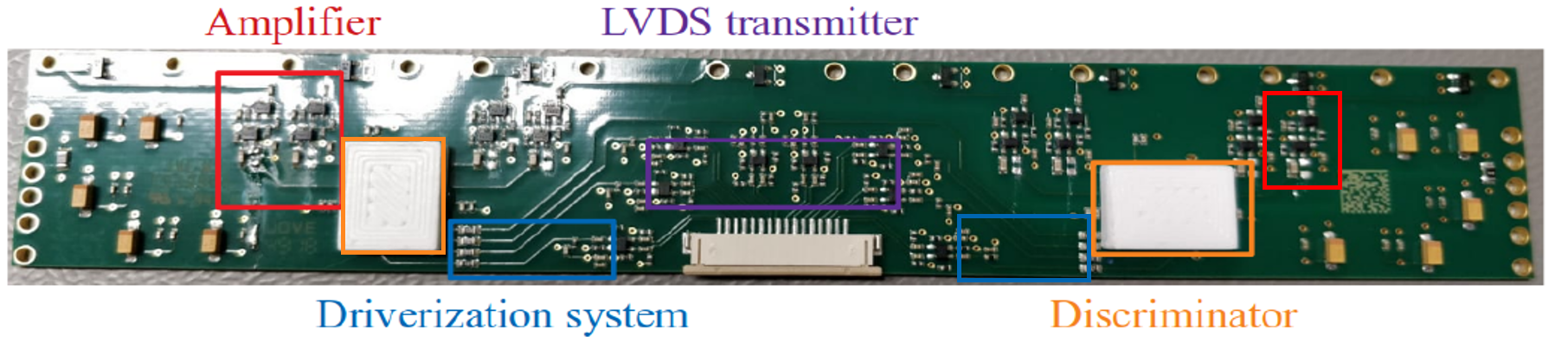}
\caption{Picture of the Front End electronics board of the BIS78 RPCs. \newline Copyright 2020 CERN for the benefit of the ATLAS Collaboration. CC-BY-4.0 license.}

\label{fesketch}
\end{figure}
\end{center}

The developed Front End electronics board, reported in Figure \ref{fesketch}, is composed of eight channels of a new preamplifier coupled with two full-custom ASIC discriminators with four channels each, and LVDS transmitters integrated directly inside the board. The overall features of the preamplifier and of the ASIC discriminator are reported in the Table \ref{FEproperties}.
\begin{table}[h!]

  \begin{center}
    \caption{Overall properties of preamplifier and full-custom ASIC discriminator}
    
    \begin{tabular}{|l|l|}
    \hline
    \multicolumn{2}{|l|}{\textbf{Preamplifier Properties}} \\
     \hline
      Si standard component&\\
      \hline
      Preamplifier sensitivity & 0.2-0.4 mV/fC  \\
 Power Consumption &3-5 V   0.5–1 mA\\
 Bandwidth & 100 MHz\\
  \hline
  
    \end{tabular}
    \begin{tabular}{|l|l|}
    \hline
    \multicolumn{2}{|l|}{\textbf{Discriminator Properties}} \\
      \hline
      SiGe full custom&\\
      \hline
Discrimination Threshold &0.5 mV\\
Power Consumption & 2-3 V   4-5 mA \\
 Bandwidth & 100 MHz\\
  \hline
  
    \end{tabular}
\label{FEproperties}  
  \end{center}
\end{table}

\section{BIS78 RPC performance with cosmic rays}
\begin{figure}[htbp]
\centering
\includegraphics[scale=0.4]{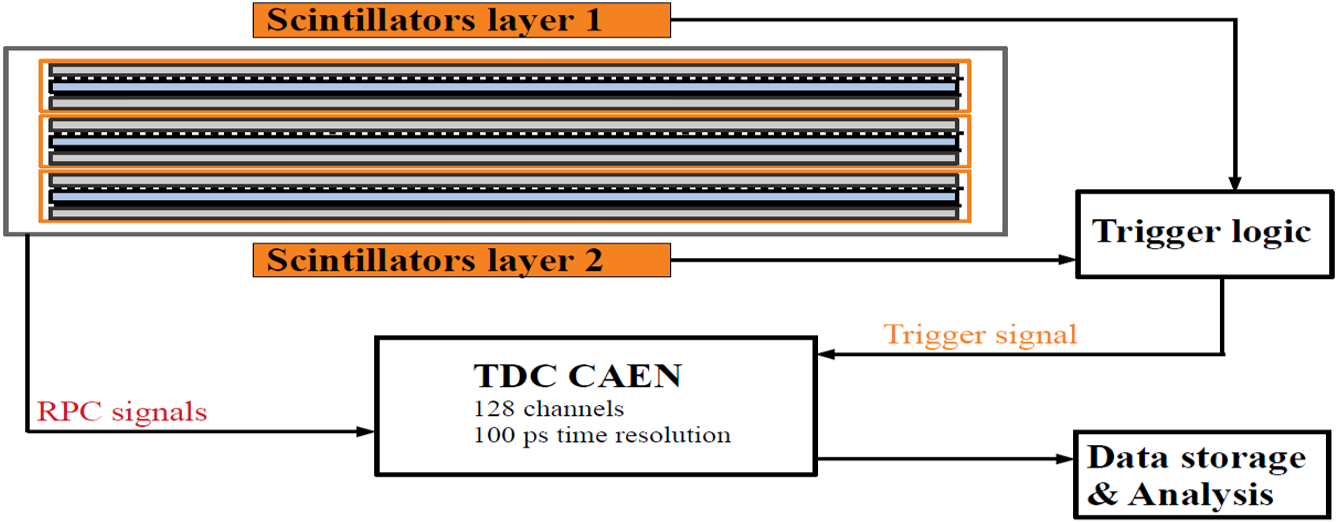}
\caption{Sketch of the experimental setup used to test the BIS78 RPCs with cosmic rays. \newline Copyright 2020 CERN for the benefit of the ATLAS Collaboration. CC-BY-4.0 license.}

\label{setup1}
\end{figure}

The BIS78 RPC performance has been tested with cosmic rays. The detector tested is composed of 3 independent mono-gap singlet equipped with the new Front-End electronics. The gas mixture used is $95 \%$ TFE, $4.7\%$  I-$C_{4}H_{10}$, $0.3\%$ SF6. A trigger system has been realized by the usage of two layers of scintillators, and the tested chamber has been put between them. The data acquisition has been realized by using a CAEN TDC V1190A with 100 ps time resolution. In Figure \ref{setup1} a scheme of the experimental setup is shown.

The main parameters monitored in order to study the detector performance with cosmic rays are the efficiency, the cluster size, the channels noise and the time resolution.
\subsection*{Efficiency}
\begin{figure}[htbp]
\centering
\includegraphics[scale=0.45]{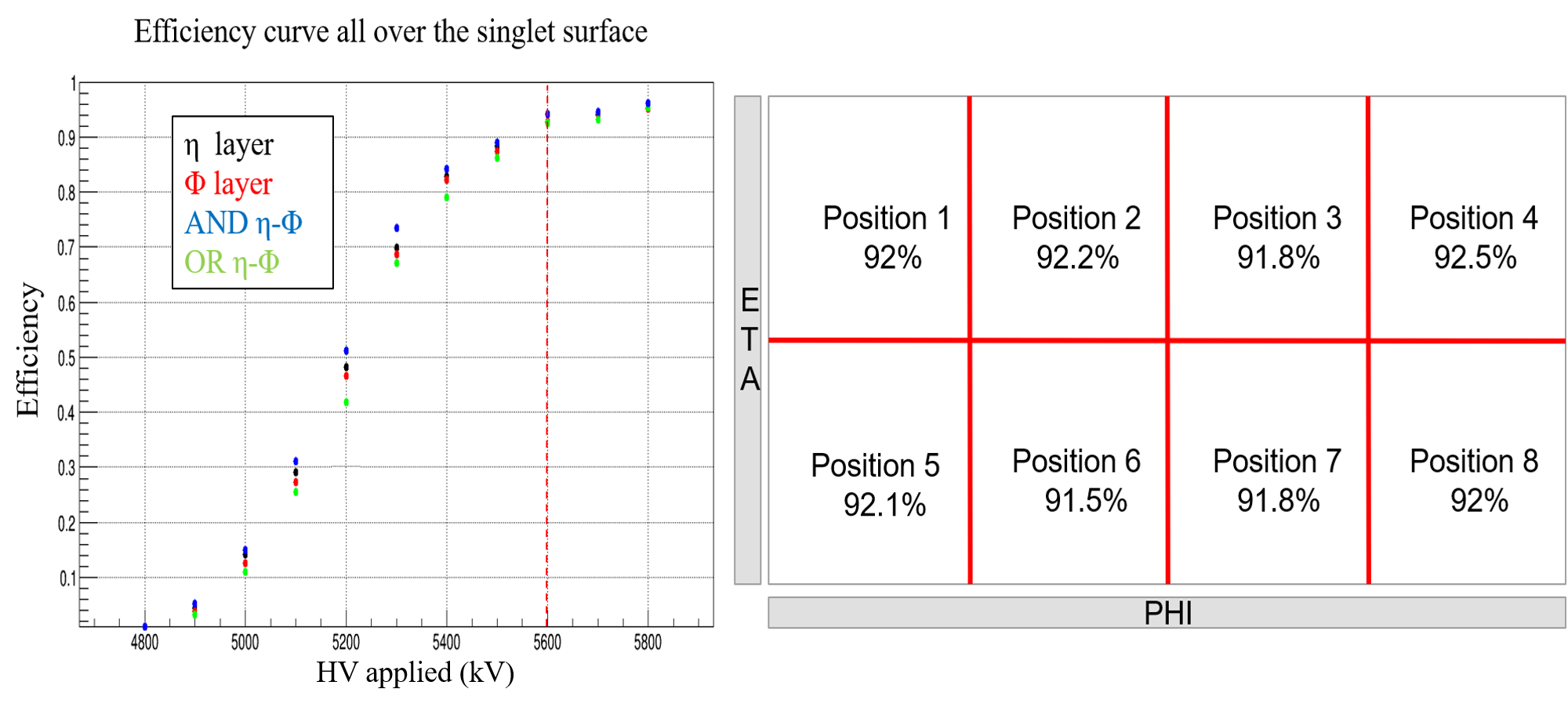}
\caption{a) Efficiency of the RPC singlet as a function of the applied High Voltage. Black and Red markers are related to both the readout panels $\eta$ and $\phi$, while the Blue and the Green ones are the logic AND and OR between them. b) Map of the efficiency all over the singlet surface estimated at the knee point (5.5kV). \newline Copyright 2020 CERN for the benefit of the ATLAS Collaboration. CC-BY-4.0 license.}

\label{eff3}
\end{figure}

The efficiency parameter has been studied in order to verify the overall behaviour of the detector. Two measurements have been performed checking this parameter: the efficiency value as a function of the applied High Voltage and the efficiency map at the knee point. In Figure \ref{eff3} the efficiency of a singlet of the BIS78 RPC system is shown.

The Figure \ref{eff3}a shows the singlet efficiency measurements as a function of HV; the red dashed line marks the $95\%$ value at 5.6 kV

 The Figure \ref{eff3}b represents the efficiency map taken at the knee point of the efficiency curve (5.5 kV), in order to be more sensitive to any efficiency variation. This map has been obtained by dividing the RPC in 8 different zones (positions) and for each zone the efficiency has been measured. The efficiency variations reported in the map are within the statistical fluctuations, hence the efficiency is homogeneous through all the detector surface. 

\subsection*{Cluster Size}
\begin{figure}[htbp]
\centering
\includegraphics[scale=0.34]{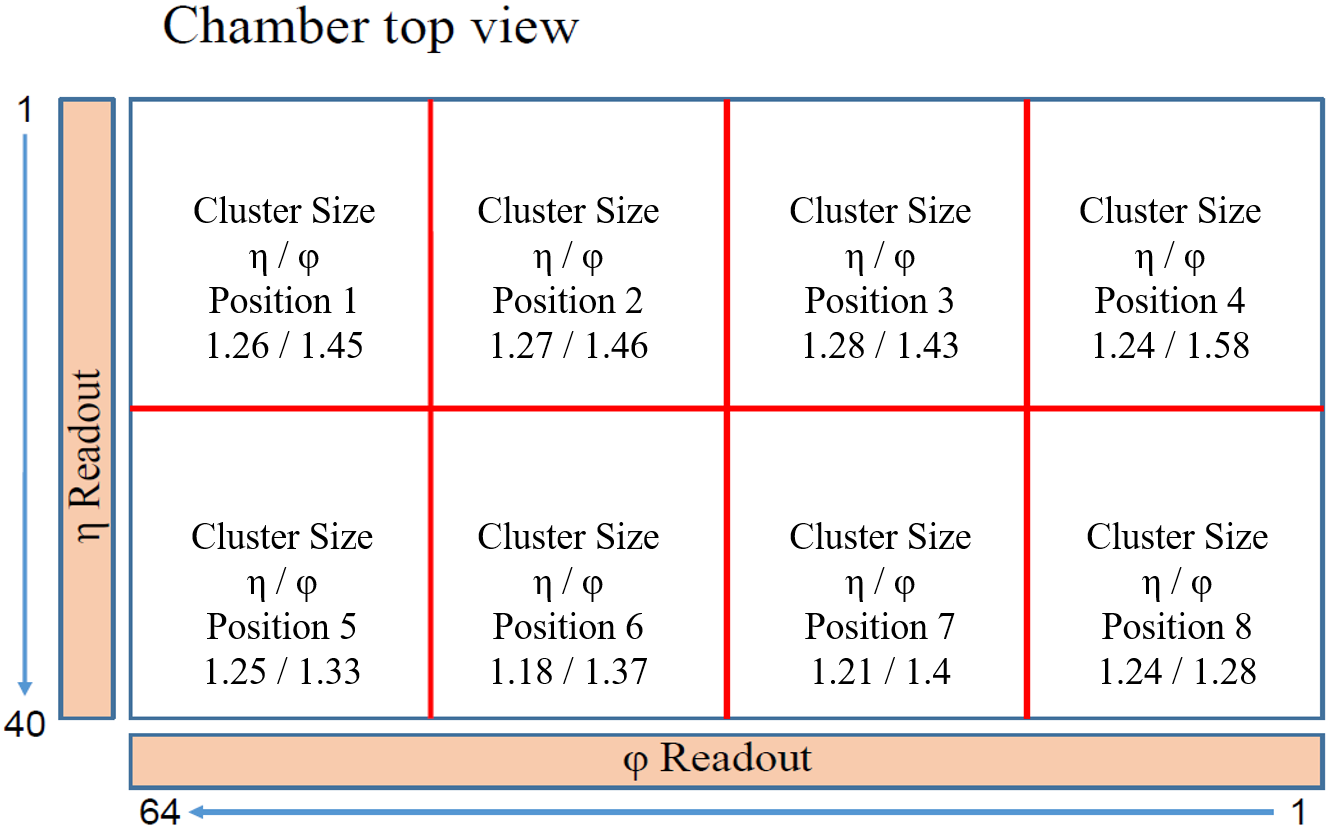}
\caption{Map of the cluster size all over the detector surface at 5.6 kV of applied High Voltage. \newline Copyright 2020 CERN for the benefit of the ATLAS Collaboration. CC-BY-4.0 license.}

\label{cs1}
\end{figure}
The Cluster Size parameter has been studied mainly to monitor the self-induced noise and the crosstalk effects, since those effects would increase the cluster size to anomalous values. The cluster size mapping has been obtained by dividing the chamber in 8 positions and for each position the size of clusters has been monitored. This map allows to verify the absence of any hot spot and to check the homogeneity of this variable. The cluster size map is reported in Figure \ref{cs1}. 

The map shows an homogeneous behaviour of the cluster size all over the detector surface with a value completely consistent with the charge induction phenomenon and a readout strip pitch of $2.5$ cm, implying that there are neither crosstalk nor self-induced noise effects.
\subsection*{Channels noise}

\begin{figure}[htbp]
\centering
\includegraphics[scale=0.27]{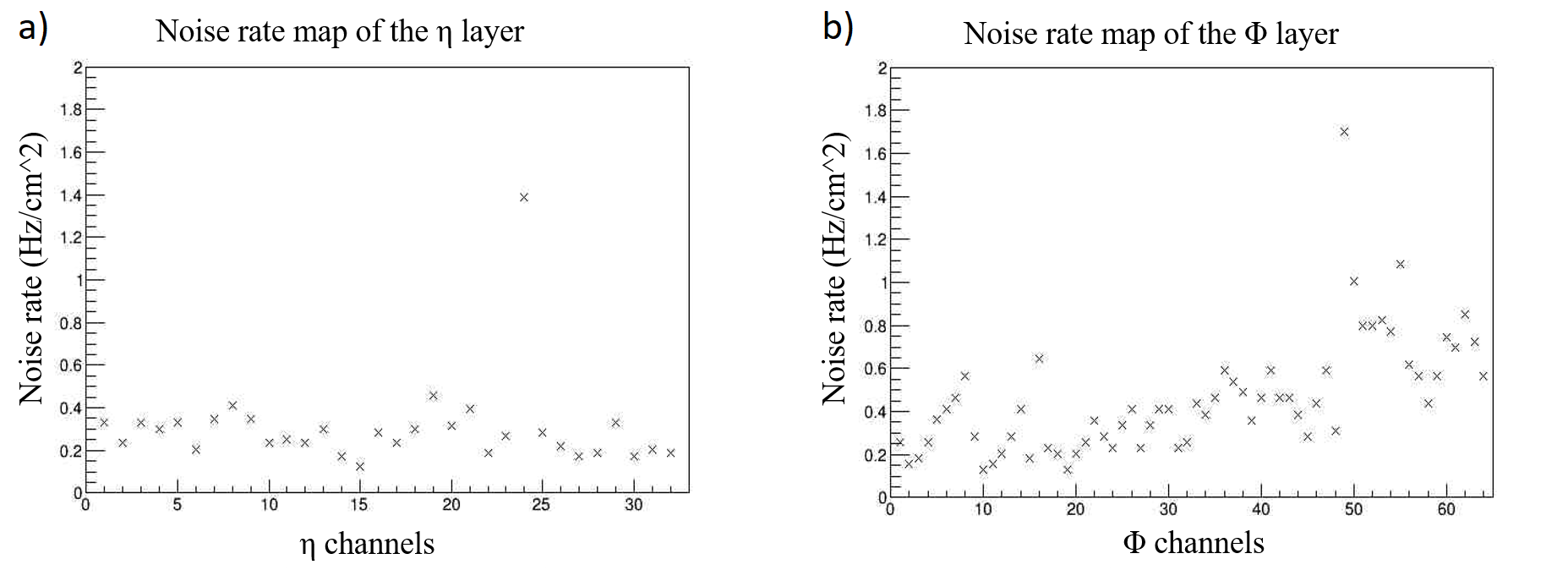}
\caption{a) Map of the noise rate of the $\eta$ layer at the working point (5.6 kV). b) Map of the noise rate of the $\phi$ layer at the working point (5.6 kV). \newline Copyright 2020 CERN for the benefit of the ATLAS Collaboration. CC-BY-4.0 license.}
\label{noise}
\end{figure}

The noise map allows to highlight problems related to disturbance and broken channels. This measurement has been performed by using a random trigger and by counting for each channel the spurious events. The channels counters have been renormalized according to the whole time acquisition window and to the strip area. The noise map reported in Figure \ref{noise} shows the noise rate in units of $Hz/cm^{2}$ of both the readout layers $\eta$ and $\phi$. The $\eta$ readout has an almost homogeneous behaviour with an average noise of 0.3 $Hz/cm^{2}$. The $\phi$ layer has an average noise rate of 0.5 $Hz/cm^{2}$. The $\phi$ layer is more susceptible to the Low Voltage connection point, which is located near the channel 64, exhibiting an higher noise rate near this region.
 
However the overall noise rate for both readout layers is within the requirements being below 1 $Hz/cm^{2}$.

\subsection*{Time resolution}

\begin{figure}[htbp]
\centering
\includegraphics[scale=0.27]{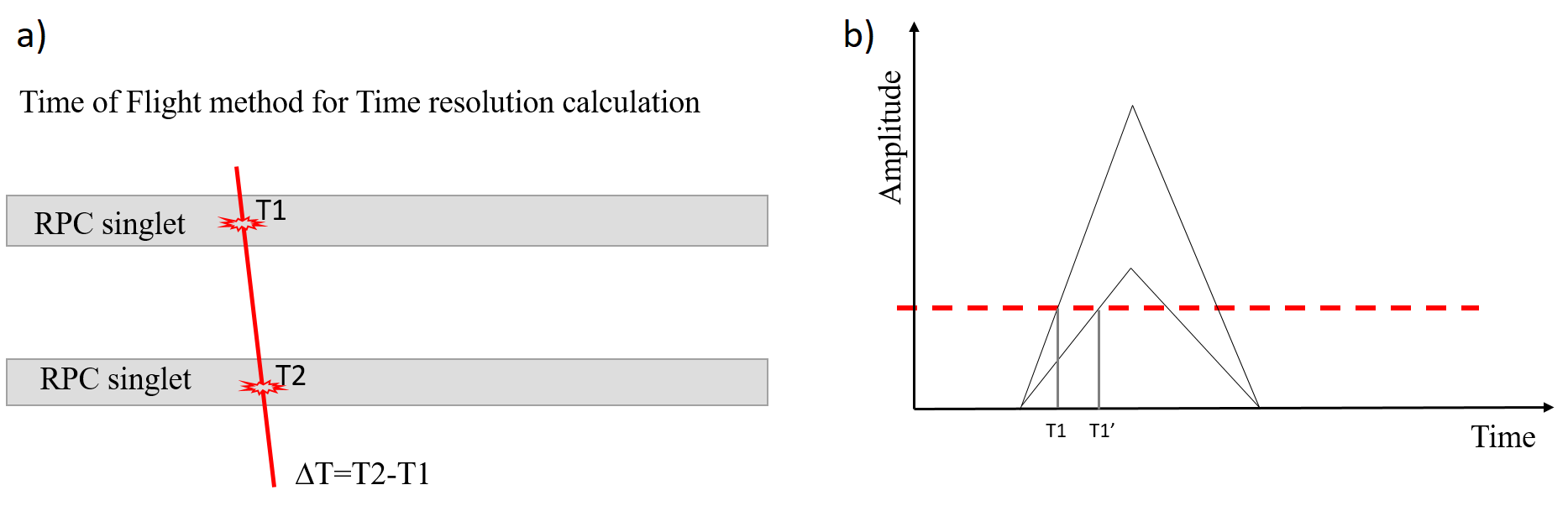}
\caption{a) Sketch of the Time-Of-Flight method. b) Sketch of the discriminator timewalk effect. \newline Copyright 2020 CERN for the benefit of the ATLAS Collaboration. CC-BY-4.0 license.}
\label{tof1}
\end{figure}

\begin{figure}[htbp]
\centering
\includegraphics[scale=0.25]{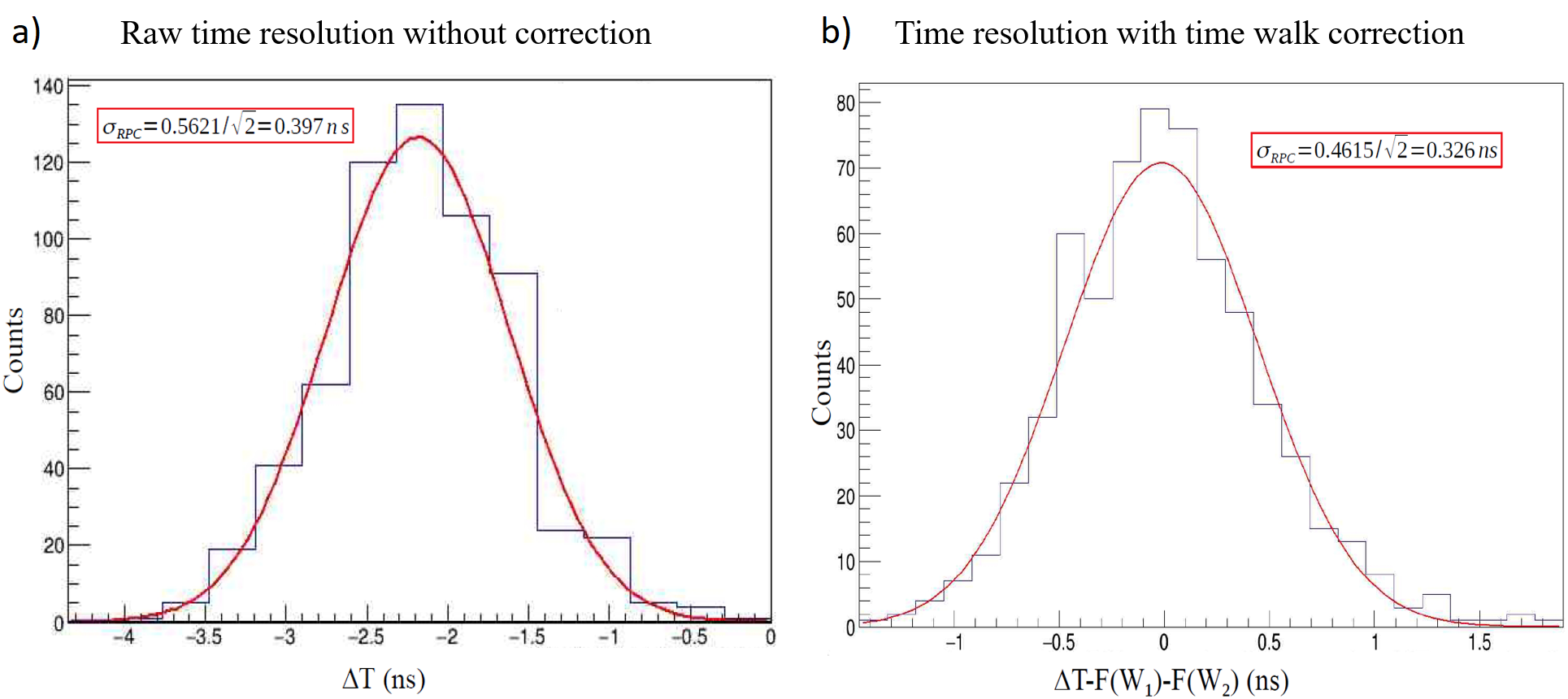}
\caption{a) Distribution of $\Delta T$ without the timewalk correction at the detector working point of 5.6 kV. b) Distribution of $\Delta T$ once the timewalk correction is applied (5.6 kV). \newline Copyright 2020 CERN for the benefit of the ATLAS Collaboration. CC-BY-4.0 license.}
\label{timeres1}
\end{figure}

The time resolution of the BIS78 RPC detectors has been tested by using the Time-Of-Flight method between 2 singlets out of 3 (all the combinations have been checked leading to the same result) and by correcting the discriminator timewalk effect with the Time-Over-Threshold measurement perfomed by the Front-End electronics itself \cite{pizzimento} (see Figure \ref{tof1}). Moreover, for each readout channel all the systematic effects have been taken into account and corrected for. In order to correct the timewalk effect, the function which correlates the amplitude of the signals with the time in which the signal surpasses the threshold has been found and corrected for event by event. In Figure \ref{timeres1} the time resolution of the BIS78 RPC is reported with and without the time walk correction, achieved at the working point (5.6 kV). 

The raw BIS78 time resolution is $\sim 400$ ps, while applying the timewalk correction leads to a slight improvement obtaining a record time resolution of $\sim 330$ ps with a monogap RPC. The improvement achieved with the timewalk correction is small due to the already very fast Front-End electronics that makes the timewalk very small \cite{pizzimento}.

\section{BIS78 RPC performance in high radiation environment}
The performance of the RPC detector equipped with the newly developed Front-End electronics under high irradiation has been the crucial point of the GIF++ testbeam at CERN \cite{H4}. Moreover, this measurement has been fundamental in order to verify the improvement in the rate capability achieved by the RPC detector by means of this new FE electronics. However the FE version used in this test was not the final one and showed some instabilities in terms of crosstalk and self induced noise, causing multiple channels to fire simulataneously or to fire multiple times on the same signal or increasing the discriminated signal width. Those problems have been solved with the final version of the FE.

This test has been carried on by setting two different sets of FE electronics parameters. One set is the most performing in terms of efficiency and minimum achievable threshold, lacking, however, in system stability. The other set is more conservative in terms of minimum achievable threshold granting on the other hand high system stability.

\begin{figure}[htbp]
\centering
\includegraphics[scale=0.5]{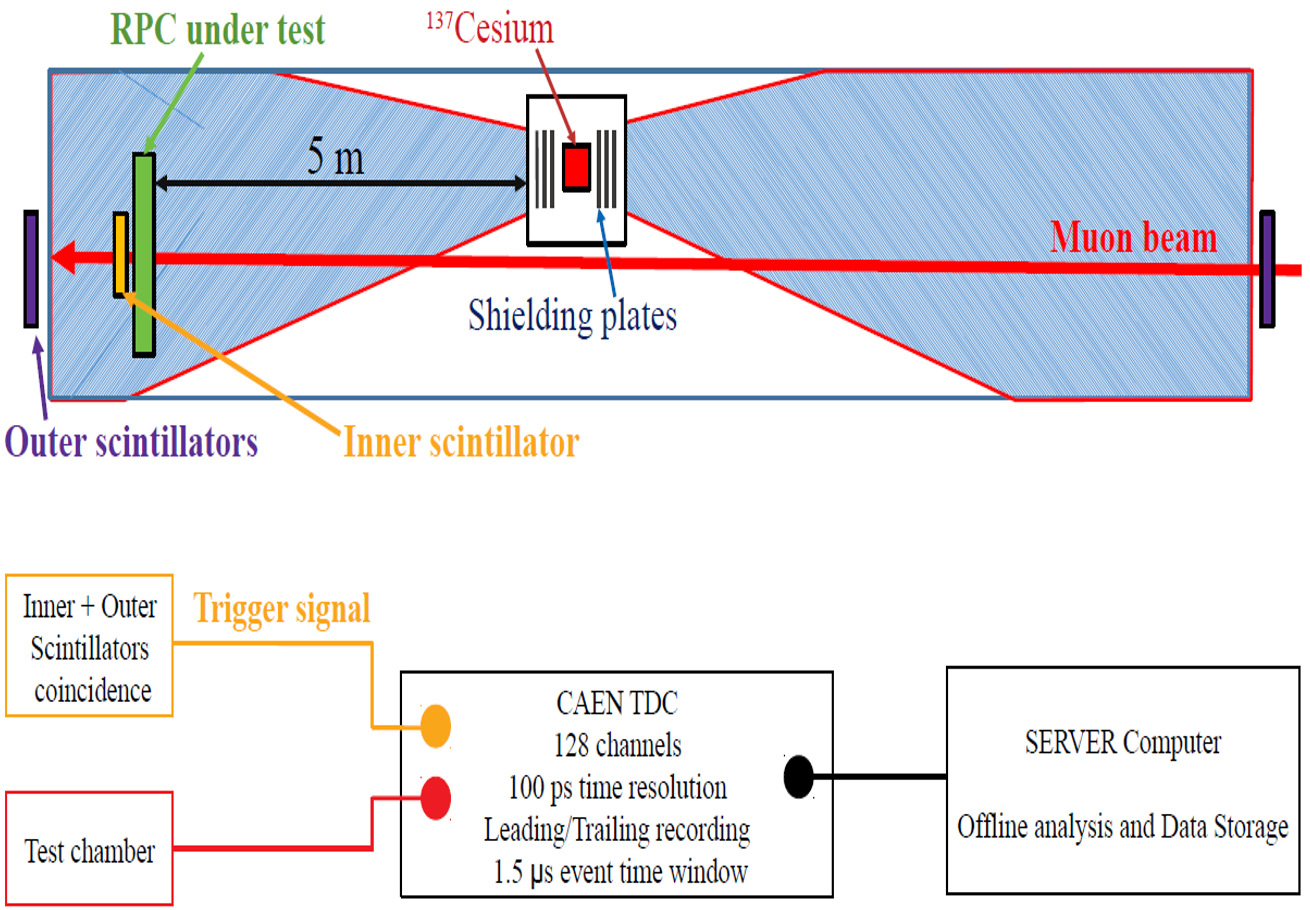}
\caption{Sketch of the experimental setup used at the GIF++ at CERN to test the BIS78 RPC detector under heavy irradiation. \newline Copyright 2020 CERN for the benefit of the ATLAS Collaboration. CC-BY-4.0 license.}
\label{setup2}
\end{figure}

The goal of this measurement was the study of the BIS78 RPC performance under high irradiation; on one hand testing the detector standard working condition when the conservative set is used and on the other hand checking the highest performance achievable when the performing parameters are set. The results achieved by both FE settings and by irradiating the detector with the $14$ $TBq$ $^{137}Cs$ source with various shielding factors and by means of a muon beam are reported. The detector has been positioned at $\sim5$ m from the source, its area was 1.8x1.1 $m^{2}$ and the whole experimental setup is reported in Figure \ref{setup2}.  

\begin{figure}[!htb]
\centering
\includegraphics[scale=0.7]{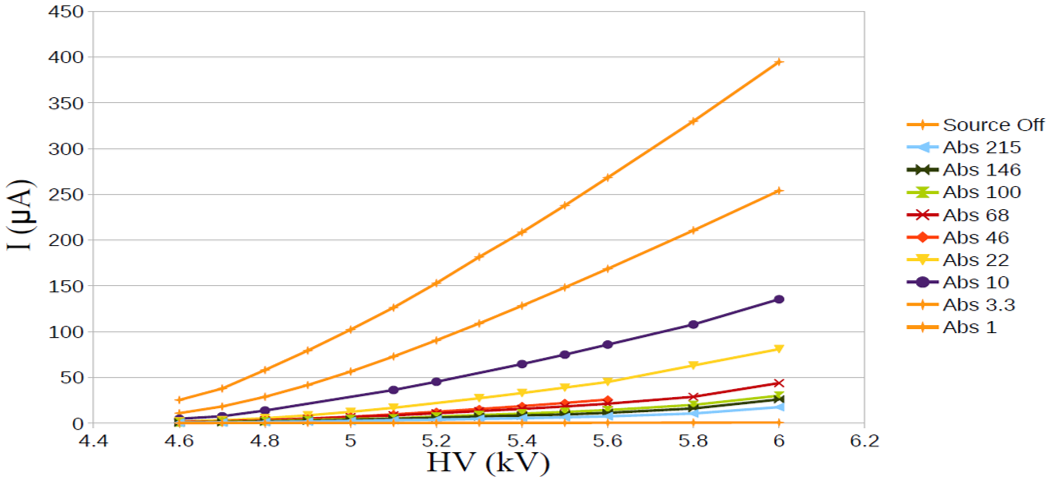} 
\caption{I-V curves of the BIS78 RPC singlet as a function of the absorption factor. \newline Copyright 2020 CERN for the benefit of the ATLAS Collaboration. CC-BY-4.0 license.}
\label{current1}
\end{figure}

The singlet I-V curve as function of the absorption factor has been monitored and it is shown in Figure \ref{current1}.

The measured current can be put in relation to the average charge per count $\braket Q$ and the counting rate $C_{photons}$ on the detector as:
\begin{equation}
I=\langle Q\rangle  \times C_{photons}
\label{formulac}
\end{equation}
The current $I$ is normalized with respect to the detector active area, hence the detector current density ($\mu A/cm^{2}$) is considered. Therefore, measuring the rate of converted photons $C_{photons}$ and the detector current with the various source shielding, the average charge $\langle Q\rangle$ per converted photon can be estimated. Since $C_{photons}$ is also the number of photons measured by the FE-RPC system, the threshold applied by the Front-End electronics can be expressed in terms of the average charge per count delivered inside the detector.

This measurement has been perfomed, considering the current density in the first higher HV point after the knee of the efficiency plateau. The Front-End thresholds applied on the charge per count are evaluated by means of the Equation \ref{formulac}. 

This study has been performed on both $\eta$ and $\phi$ readouts for different absorption factors \cite{pizzimento} and for both performing and conservative threshold settings.

\begin{figure}[htbp]
\centering
\includegraphics[scale=0.4]{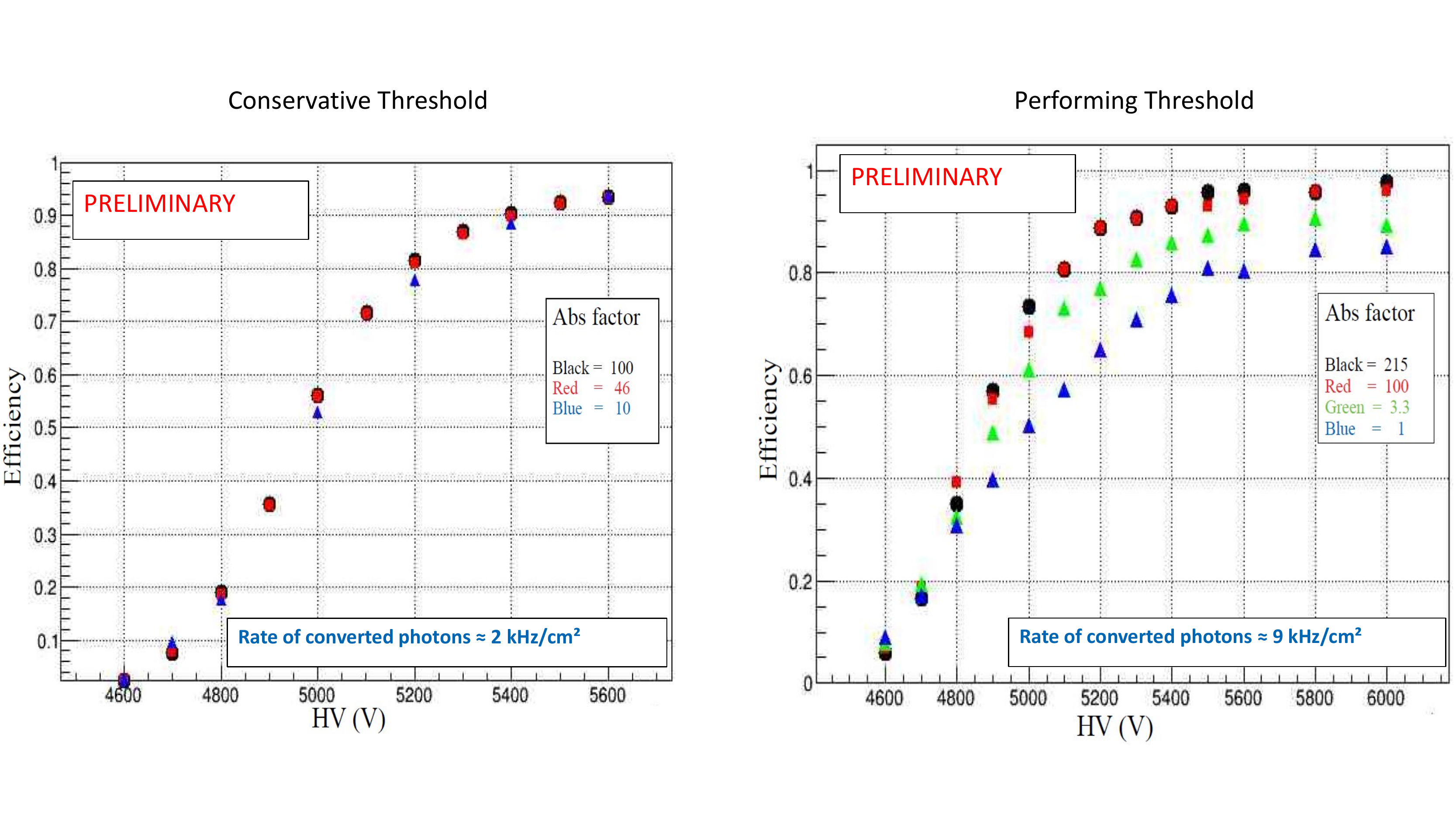}
\caption{a) Efficiency curves of the RPC detector when the conservative threshold is used, varying the source absorption factor. b) Efficiency curves of the RPC detector when the performing threshold is set and for different source absorption factor. \newline Copyright 2020 CERN for the benefit of the ATLAS Collaboration. CC-BY-4.0 license.}
\label{rate}
\end{figure}

By means of the FE conservative setting an average charge per count produced inside the gas $\braket Q$ of $5\pm 1$ pC was achieved.

The efficiency curves for muons with the conservative settings and for different absorption factors are reported in Figure \ref{rate}a.
The efficiency curves do not show any deviation from the source-off behaviour. Operating the detector in these conditions leads to no efficiency losses for rate of converted photons up to 2 $kHz/cm^{2}$. 

Taking into account the FE performing setting, the estimated FE threshold in terms of average charge per count on both $\eta$ and $\phi$ readouts is $\sim 3 \pm 1$ pC. The big uncertainty is due to the difficulty to correctly estimate the occupancy and the measured rate, due to system instabilities described before.

The efficiency curves achieved with the perfoming threshold and with different absorption factors are reported in Figure \ref{rate}b. It must be underlined how those efficiency values, especially for very low absorption factors ($<10 Abs$), represent only the lower limit to the real detector efficiency which is higher, since the high level of occupancy is the main cause for the flat efficiency losses. However with the performing threshold set a rate capability up to 9 $kHz/cm^{2}$ has been achieved, confirming how the Front-End electronics achieved a $\sim3$ pC charge threshold on the average charge per count produced within the gas gap.

\section{Conclusion}
The BIS78 project is the pilot project for the ATLAS detector upgrade of the barrel muon spectrometer foreseen to meet the requirements of the HL-LHC running conditions. The BIS78 RPCs detectors grant a record combined performance of better than $95\%$ single gap efficiency and time resolution of $\sim330$ $ps$. The highest rate capability achieved by this system is $9$ $kHz/cm^{2}$ with an efficiency above $80\%$. Further tests are ongoing in order to fully demonstrate that the new version of FE addresses the instability problems leading to a fully efficient detector with such high rate capability. Moreover, the succesfulness of the integration of the FE electronics within a large area detector has been demonstrated, reaching high performance along with a good homogeneity overall behaviour through all the BIS78 modules produced.

%


\end{document}